\documentclass[12pt,preprint]{emulateapj}
\usepackage{graphicx}

\shorttitle{Transit Potential of iota Draconis b}
\shortauthors{Stephen R. Kane et al.}
\slugcomment{Submitted for publication in the Astrophysical Journal}
\begin{document}

\title{On the Transit Potential of the Planet Orbiting iota Draconis}

\author{
Stephen R. Kane\altaffilmark{1},
Sabine Reffert\altaffilmark{2},
Gregory W. Henry\altaffilmark{3},
Debra Fischer\altaffilmark{4},
Christian Schwab\altaffilmark{2},
Kelsey I. Clubb\altaffilmark{4},
Christoph Bergmann\altaffilmark{2}
}
\email{skane@ipac.caltech.edu}
\altaffiltext{1}{NASA Exoplanet Science Institute, Caltech, MS 100-22,
  770 South Wilson Avenue, Pasadena, CA 91125, USA}
\altaffiltext{2}{ZAH-Landessternwarte, K\"onigstuhl 12, 69117
  Heidelberg, Germany}
\altaffiltext{3}{Center of Excellence in Information Systems, Tennessee
  State University, 3500 John A. Merritt Blvd., Box 9501, Nashville,
  TN 37209, USA}
\altaffiltext{4}{Department of Physics \& Astronomy, San Francisco
  State University, San Francisco, CA 94132, USA}

%%%%%%%%%%%%%%%%%%%%%%%%%%%%%%%%%%%%%%%%%%%%%%%%%%%%%%%%%%%%%%%%%%%%

\begin{abstract}

Most of the known transiting exoplanets are in short-period orbits,
largely due to the bias inherent in detecting planets through the
transit technique. However, the eccentricity distribution of the known
radial velocity planets results in many of those planets having a
non-negligible transit probability. One such case is the massive planet 
orbiting the giant star iota Draconis, a situation where both the orientation 
of the planet's eccentric orbit and the size of the host star inflate 
the transit probability to a much higher value than for a typical hot 
Jupiter. Here we present a revised fit of the radial velocity data with 
new measurements and a photometric analysis of the stellar variability. We
provide a revised transit probability, an improved transit ephemeris, and 
discuss the prospects for observing a transit of this planet from both 
the ground and space.

\end{abstract}


%%%%%%%%%%%%%%%%%%%%%%%%%%%%%%%%%%%%%%%%%%%%%%%%%%%%%%%%%%%%%%%%%%%%

\section{Introduction}
\label{introduction}

The detection of exoplanetary transits has revolutionized the way 
we view giant planets, both in their variety of structures
\citep{bat09} and their formation processes \citep{ver09}. Even so,
this view is mostly restricted to planets in short-period circular
orbits around their host stars. The ground-based transit detection of
the planets HD~17156b \citep{bar07a} and HD~80606b
\citep{lau09,mou09}, enabled by their high eccentricities
\citep{kan08,kan09a}, provided the first insights into the structures
of longer-period planets. Further discoveries of long-period planetary
transits around bright stars are vital to understanding the dependence
of planetary structure/evolution on the periastron distance of the
planet \citep{kan09b}. Provided the orbital parameters can be
determined with sufficient precision, monitoring planets detected
via the radial velocity technique at predicted transit times
provides a means to increase the sample of long-period transiting
planets. Efforts to detect transits of the known radial velocity
planets are currently being undertaken by the Transit Ephemeris
Refinement and Monitoring Survey (TERMS).

Amongst the radial velocity planet discoveries, a substantial number
have been found to orbit giant stars. Recent examples of such
discoveries include the planetary companions of 11 Ursae Minoris and
HD~32518 \citep{dol09} and HD~102272 \citep{nie09}. These detections
are in good agreement with predictions of their frequency and
dependence on host star mass \citep{ida05,ken08}. In particular,
planets orbiting giants stars tend to have large transit probabilities
due to the size of the host stars \citep{ass09}. Giant stars present
significant challenges, however, to those who intend to monitor those
stars for the purpose of detecting exoplanetary transits. A good
example is the planet orbiting HD~122430 \citep{set04}, for which the
combination of the 22.9 $R_\sun$ host star \citep{das06} and the high
eccentricity of the planetary orbit lead to a transit probability of
$\sim 32$\%. However, assuming a Jupiter radius for the planet
requires the unambiguous detection of a $1.9 \times 10^{-5}$ transit
depth. Additionally, the uncertainty in the orbital parameters of the
planet means that a large transit window will need to be continuously
monitored \citep{kan09b}. Finally, a photometric survey of giant stars
by \citet{hen00} found that almost half of their sample exhibited
low-level photometric variability that would further complicate
transit detection.

The planet orbiting iota Draconis (hereafter $\iota$ Dra) presents a
particularly interesting case. The host star is a K2 giant, is
very bright ($V = 3.29$), and is frequently referred to by its other
common aliases of HD~137759 and HIP~75458. A thorough spectral analysis
of this star was undertaken by \citet{sad05}, who found a metallicity
of [Fe/H] $= 0.12$. The planetary companion was discovered by
\citet{fri02}, and the orbit was further refined by \citet{zec08}, whose
radial velocity data revealed a linear trend over time. In addition to
the large stellar radius, the planetary orbit is highly eccentric and
the argument of periastron ($\omega \sim 90\degr$) ensures that the
periastron passage occurs approximately in the observer-star plane
perpendicular to the line-of-sight.

Here we present new radial velocity data for $\iota$ Dra b and an
analysis of the photometric stability of the host star. These data are
used to provide a well-constrained transit ephemeris for the next 10
years and an assessment of the feasibility of detecting a transit for
the planetary companion. This analysis and discussion may be used as a
model for how to consider the transit detection potential for each of
the planets orbiting giant stars. In Section 2 we present the new
radial velocity data and revised orbital solution. Section 3 discusses
the geometric transit probability of the planet and the predicted
transit depth, and in Section 4 we calculate the transit ephemeris for
the planet. Section 5 presents the photometric analysis of the host
star, and in Section 6 we discuss the potential for detecting a
transit of the planet.

%%%%%%%%%%%%%%%%%%%%%%%%%%%%%%%%%%%%%%%%%%%%%%%%%%%%%%%%%%%%%%%%%%%%

\section{Orbital Solution}

The initial orbital solution provided by \citet{fri02} had relatively
large uncertainties associated with the orbital parameters because
they had only a single cycle of the planetary orbit. The revised
solution by \citet{zec08} is a significant improvement, mostly due to
the large increase in time baseline leading to a much
better-constrained period. Here we combine the Lick Observatory data
published by \citet{zec08} with further data covering almost two
additional cycles of the orbit. The additional data were also acquired
with the 0.6m Coud\'e Auxillary Telescope and the Hamilton Echelle
Spectrograph at Lick Observatory with the same techniques described by
\citet{zec08}.

\begin{figure}
  \begin{center}
    \begin{tabular}{c}
      \includegraphics[angle=270,width=8.2cm]{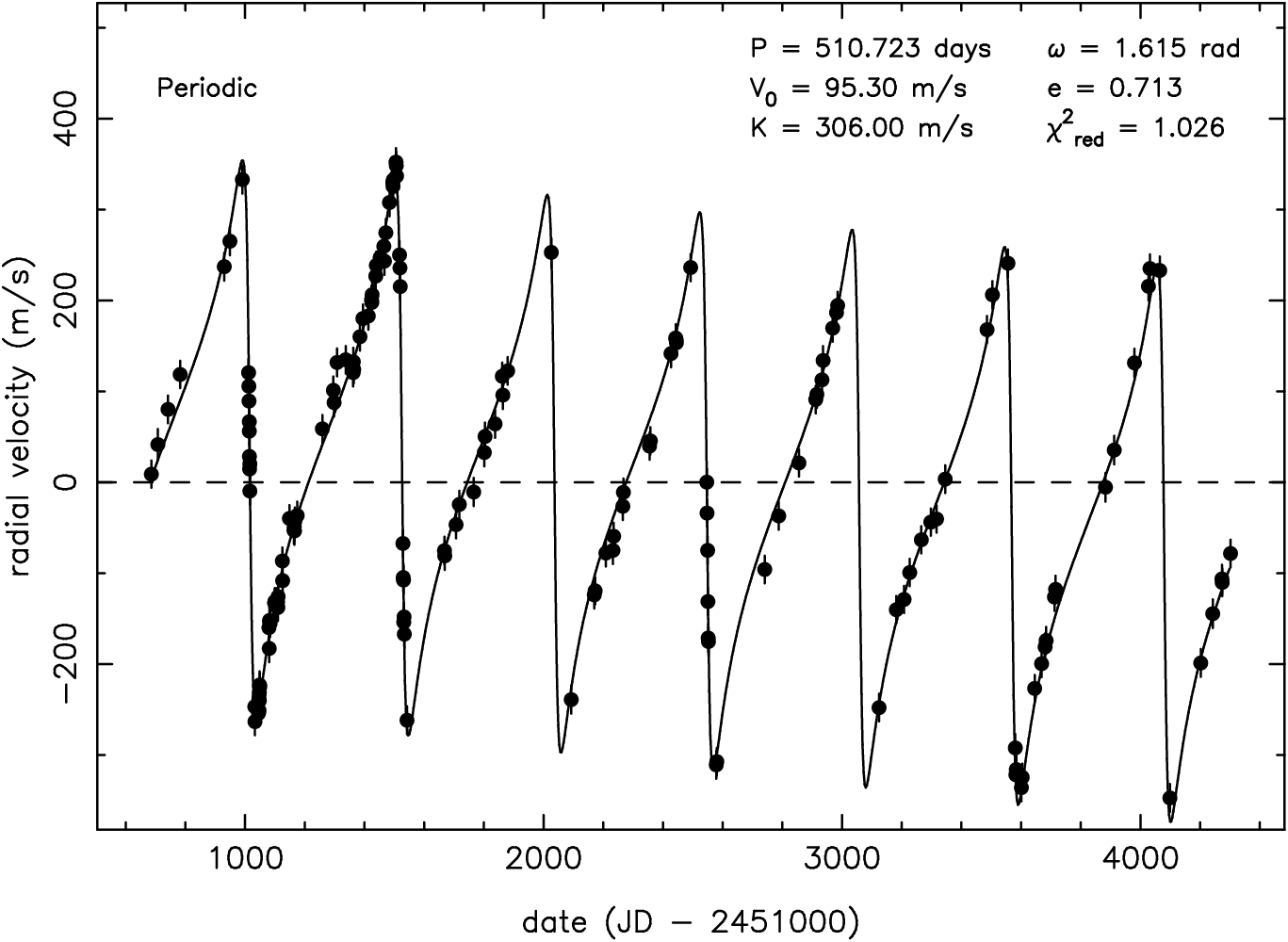} \\
      \includegraphics[angle=270,width=8.2cm]{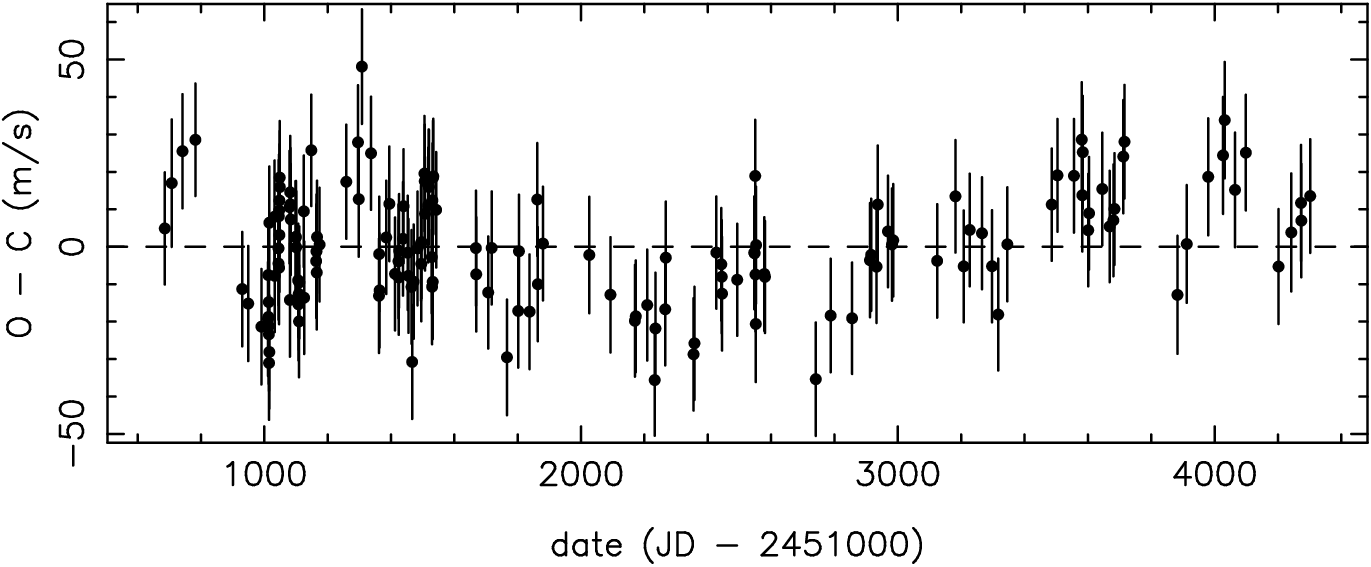}
    \end{tabular}
  \end{center}
  \caption{Radial velocity measurements of $\iota$ Dra along with
    the best-fit orbital solution (solid line). The lower panel shows
    the residuals of the fit, observed minus calculated ($O - C$).}
  \label{rvmodel}
\end{figure}

The combined data are shown in Figure \ref{rvmodel} along with the
best-fit Keplerian model. A noise component of 14.5 m/s was added in
quadrature to the derived uncertainties for each measurement in order
to force $\chi^2 \sim 1$. As was noted by \citet{zec08}, the addition
of a linear trend of $\sim -13.65$ m/s/yr is required to provide an
adequate fit to the data. The residuals to the fit, shown in the lower
panel of Figure \ref{rvmodel}, have an RMS of 15.05 m/s. This level of
stellar jitter is common for evolved stars
\citep{wri05}. \citet{zec08} find a particularly strong stellar
oscillation signature for $\iota$ Dra with a frequency of 3.8
days$^{-1}$. A fourier analysis of the residuals did not reveal any
further significant periodicity. The orbital parameters for the fit
presented here are shown in Table \ref{param}, along with the
parameters determined by \citet{zec08} for comparison.

\begin{table}
  \begin{center}
    \caption{Fit parameters for $\iota$ Dra.}
    \label{param}
    \begin{tabular}{@{}lcc}
      \hline
      Parameter & 2008 fit & Revised fit \\
      \hline
$P \ (\mathrm{days})$         & $510.88 \pm 0.15$   & $510.72 \pm 0.07$ \\
$K \ (\mathrm{km \ s^{-1}})$  & $299.9  \pm 4.3$    & $306.0  \pm 3.8$ \\
$\omega \ (\degr)$            & $88.7 \pm 1.4$      & $92.5   \pm 0.7$ \\
$e$                           & $0.7261 \pm 0.0061$ & $0.713 \pm 0.008$ \\
$t_p \ (\mathrm{JD}-2450000)$ & $2013.94 \pm 0.48$  & $2015.36 \pm 0.16$ \\
Linear trend (m/s/yr)         & $-13.8 \pm 1.1$     & $-13.65 \pm 0.75$ \\
      \hline
    \end{tabular}
    \tablecomments{The orbital parameters for $\iota$ Dra as measured
      by \citet{zec08}, and the revised orbital parameters based upon
      the combined dataset presented in this paper.}
  \end{center}
\end{table}

\begin{figure}
  \includegraphics[angle=270,width=8.2cm]{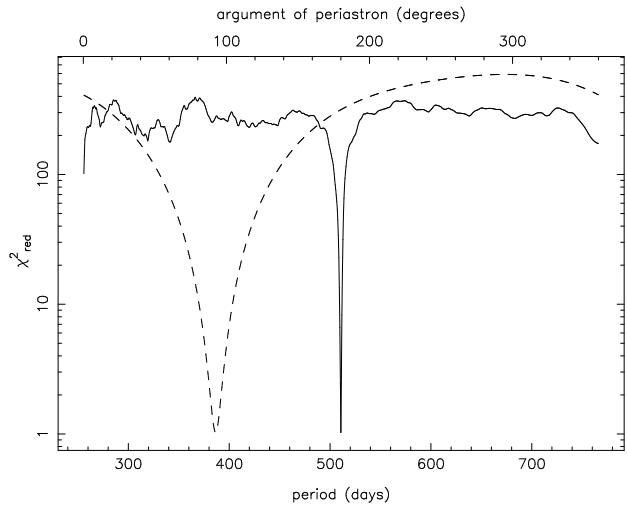}
  \caption{$\chi^2$ maps for the fitted period (solid line) and
    argument of periastron (dashed line).}
  \label{chi2maps}
\end{figure}

Although the eccentricity and periastron argument have comparable
precisions resulting from this fit, the ephemeris critical parameters
of period and time of periastron passage \citep{kan09b} are
significantly improved. This is represented in Figure \ref{chi2maps},
which shows the reduced $\chi^2$ maps of parameter space for both the
period (solid line) and the periastron argument (dashed line), in
which one parameter is varied whilst holding all other parameters
fixed. The large number of orbital phases covered prevents significant
global minima from occurring in period space, but the periastron
argument is less constrained due to the need for more measurements
during the eccentric (rapidly changing) phase of the planetary orbit.

%%%%%%%%%%%%%%%%%%%%%%%%%%%%%%%%%%%%%%%%%%%%%%%%%%%%%%%%%%%%%%%%%%%%

\section{Transit Probability, Depth, and Duration}

\begin{figure}
  \includegraphics[angle=270,width=8.2cm]{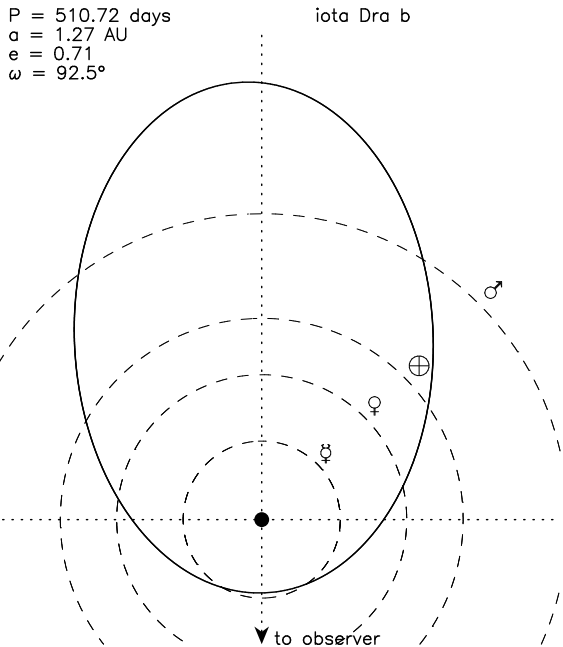}
  \caption{The orbit of the planet orbiting $\iota$ Dra (solid line)
    and the orbits of the Solar System planets for comparison (dashed
    lines).}
  \label{orbit}
\end{figure}

As described by \citet{bar07b} and \citet{kan08}, the transit
probability is a strong function of both the eccentricity and the
argument of periastron. In particular, the transit probability is the
strongest when the periastron passage occurs close to the
star-observer line of sight, or where $\omega = 90\degr$. The orbit
for $\iota$ Dra b, shown in Figure \ref{orbit}, is thus very well
suited to produce a relatively high transit probability.

\begin{figure}
  \includegraphics[angle=270,width=8.2cm]{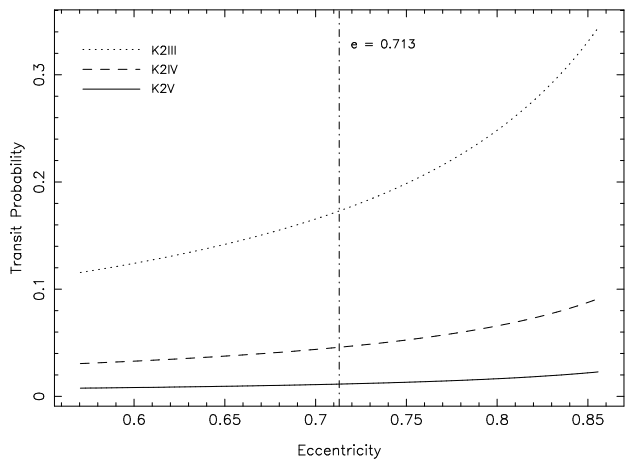}
  \caption{Transit probability as a function of eccentricity for three
    different stellar models, showing the dramatic impact of the host
    being a giant star.}
  \label{eccprob}
\end{figure}

The other two primary factors that impact the transit probability are
the radii of the planet and the host star. One can see from Figure
\ref{orbit} that $\iota$ Dra b would have a similar transit
probability to Mercury if it wasn't for these additional factors. To
demonstrate the impact of the stellar radius and orbital eccentricity,
Figure \ref{eccprob} shows how the transit probability varies as a
function of eccentricity for three different luminosity classes for a 
K2 star. This uses the best-fit orbital parameters shown in
Table \ref{param} and assumes a planetary radius of one Jupiter
radius. The effect of eccentricity is clearly dramatic for giant
stars, driving the transit probability past 30\% for eccentricities
greater than $\sim 0.85$.

However, the dramatic positive effect the stellar radius has upon the
transit probability results in an equally dramatic negative effect
upon the transit depth. For the host star mass and radius, we adopt
the values of $M_\star = 1.05 M_\odot$ and $R_\star = 12.88 R_\odot$
respectively, as measured by \citet{all99}. This stellar mass produces
a planetary mass of $M_p \sin i = 8.7 M_J$ and semi-major axis of $a =
1.27 \ \mathrm{AU}$. Using the methods described in \citet{bod03}, we
estimate a planetary radius of $R_p = 1.11 R_J$. The uncertainty in
the stellar mass/radius and subsequent uncertainty in the planetary
mass/radius has a minor effect on the estimated transit duration but
in no way effects the predicted transit mid-points since these are
derived using the orbital parameters. Additionally, planetary radii
become insenstive to mass in the high-mass regime, as demonstrated by
the high-mass planets XO-3~b and CoRoT-3~b. Planetary radii have very
little impact on both the transit probability and duration since $R_p
\ll R_\star$, but the insensitivity of radius to mass as described
makes this impact neglible in nature.

Table \ref{probdepdur} shows the calculated transit probability
($P_t$), depth, and duration ($t_d$) for $\iota$ Dra b using the
orbital parameters from Table \ref{param}. For comparative purposes,
these parameters are also shown for a solar radius star. The problem
of the transit depth becomes very apparent when one draws this
comparison and, as described earlier, is one of the main hinderances
to ground-based detection of such transits. The expected long duration
of the transit is also problematic since ground-based coverage of a
transit from a single telescope is impossible. These issues are
described in more detail in the following sections.

\begin{table}
  \begin{center}
    \caption{Transit probabilities, depths, and durations.}
    \label{probdepdur}
    \begin{tabular}{@{}ccccc}
      \hline
      $R_\star$ & $e$ & $P_t$ (\%) & $\Delta F / F_0$ & $t_d$ (days)\\
      \hline
      1.00  & 0.000 &  0.41 & $1.30 \times 10^{-2}$ & 0.63 \\
      1.00  & 0.713 &  1.42 & $1.30 \times 10^{-2}$ & 0.26 \\
      12.88 & 0.000 &  4.74 & $7.84 \times 10^{-5}$ & 7.42 \\
      12.88 & 0.713 & 16.52 & $7.84 \times 10^{-5}$ & 3.04 \\
      \hline
    \end{tabular}
     \tablecomments{These calculations assume a planetary radius of
       $R_p = 1.11 R_J$.}
  \end{center}
\end{table}

%%%%%%%%%%%%%%%%%%%%%%%%%%%%%%%%%%%%%%%%%%%%%%%%%%%%%%%%%%%%%%%%%%%%

\section{Transit Ephemeris Refinement}

The accuracy of the predicted transit mid-point, and the size of the
time window in which the complete transit could occur, are of course
highly dependent upon the uncertainties associated with the orbital
parameters. The calculations for predicting transit ephemerides and
the influence of period and time of periastron passage are described
in detail by \citet{kan09b}. Based upon the improved orbital
parameters shown in Table \ref{param}, the transit ephemeris for
$\iota$ Dra b has been calculated for the next 10 years and is shown
in Table \ref{ephem}.

\begin{table*}
  \begin{center}
    \caption{Refined transit ephemeris for $\iota$ Dra b.}
    \label{ephem}
    \begin{tabular}{@{}|c|c|c|c|c|c|}
      \hline
      \multicolumn{2}{|c|}{Beginning} &
      \multicolumn{2}{|c|}{Mid-point} &
      \multicolumn{2}{|c|}{End} \\
      \hline
      JD & Date & JD & Date & JD & Date \\
      \hline
      2455587.81 & 2011 01 26 07 20 & 2455589.98 & 2011 01 28 11 35 & 2455592.16 & 2011 01 30 15 50\\
      2456098.46 & 2012 06 19 22 56 & 2456100.70 & 2012 06 22 04 52 & 2456102.95 & 2012 06 24 10 48\\
      2456609.11 & 2013 11 12 14 32 & 2456611.42 & 2013 11 14 22 09 & 2456613.74 & 2013 11 17 05 46\\
      2457119.76 & 2015 04 07 06 08 & 2457122.14 & 2015 04 09 15 26 & 2457124.53 & 2015 04 12 00 43\\
      2457630.41 & 2016 08 29 21 44 & 2457632.86 & 2016 09 01 08 43 & 2457635.32 & 2016 09 03 19 41\\
      2458141.06 & 2018 01 22 13 20 & 2458143.58 & 2018 01 25 01 59 & 2458146.11 & 2018 01 27 14 38\\
      2458651.71 & 2019 06 17 04 56 & 2458654.30 & 2019 06 19 19 16 & 2458656.90 & 2019 06 22 09 36\\
      2459162.36 & 2020 11 08 20 32 & 2459165.02 & 2020 11 11 12 33 & 2459167.69 & 2020 11 14 04 34\\
      \hline
    \end{tabular}
    \tablecomments{The columns indicate the beginning, mid-point, and
      end of the transit window in both Julian and calendar date for
      the next 10 years. The calendar date is expressed in UT and
      includes the year, month, day, hour, and minute.}
  \end{center}
\end{table*}

The size of the new predicted transit window is 73.56 days based upon
the orbital parameters of \citet{fri02} and 6.05 days based upon the
orbital parameters of \citet{zec08}. With the improved orbital
parameters described in this paper, the next predicted transit window
has been reduced in size to 4.35 days. By the time the 2020 transit
prediction arrives, the size of the transit window will have grown to
5.33 days. It should be noted that, despite the remaining large size
of the transit window, it is now largely dominated by the transit
duration, which is expected to be $\sim 3.0$ days as shown in Table
\ref{probdepdur}. Even so, attempts to obtain full coverage of the transit
window from the ground will require a multi-longitudinal campaign
during which one can only hope for cooperative weather. The complete
observation of an ingress or egress during a single night is a
substantially more achievable goal under such circumstances. However,
one still must contend with the challenge of meeting the photometric
precision requirements for a successful detection.

%%%%%%%%%%%%%%%%%%%%%%%%%%%%%%%%%%%%%%%%%%%%%%%%%%%%%%%%%%%%%%%%%%%%

\section{Photometric Stability}

SIMBAD refers to $\iota$~Dra as a variable star based on its citation
as NSV~7077 in the {\it New Catalogue of Suspected Variable Stars}
\citep{kuk82}. The NSV entry is based on the photometric study of
\citet{jac63}, who reported a magnitude range of 0.09 mag. Later,
\citet{per93} included $\iota$~Dra in his search for photometric
variability in K giants chosen from the {\it Bright Star Catalogue}
\citep{hof91} and found the star to be constant to a limit of 0.01
mag.

\begin{figure}
  \includegraphics[angle=270,width=8.2cm]{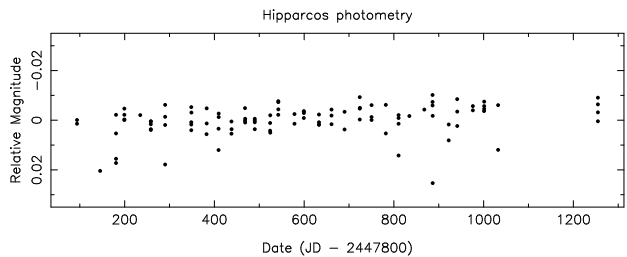}
  \caption{Photometry of $\iota$ Dra from the Hipparcos mission.}
  \label{phot_hip}
\end{figure}

We investigated the photometric stability of $\iota$~Dra using newer
observations. The {\it Hipparcos} satellite observed the star during
its three-year mission and acquired a photometric data set consisting
of 104 measurements spanning a period of 1160 days \citep{per97}. The
scatter of the 104 $\iota$~Dra measurements is 0.005 mag, while the
range of the observations, defined in terms of the 5th and 95th
percentiles of their distribution, is 0.02 mag. The scatter is roughly
consistent with the expected uncertainty of a single observation, but
the range is roughly twice that expected from a constant
star. Consequently, the {\it Hipparcos} Catalogue \citep{per97} lists
the variability type for $\iota$~Dra as a blank, indicating that the
star ``could not be classified as variable or constant.''  We
performed a Fourier analysis of the {\it Hipparcos} data, plotted in
Figure \ref{phot_hip}, and confirmed the absence of any significant
periodic variability.

We also acquired new Johnson $B$ and $V$ observations with the T3
0.4~m Automatic Photoelectric Telescope (APT) located at Fairborn
Observatory in the Patagonia Mountains of southern Arizona. Between
2010 January and May, T3 observed $\iota$~Dra differentially with
respect to a nearby comparison star in the following sequence, termed
a group observation: {\it S,C,V,C,V,C,V,C,S}, where $S$ is a sky
reading, $C$ is the comparison star HD 144284 = $\theta$~Dra ($V=4.01$,
$B-V=0.53$, F8~IV), and $V$ is the program (variable?) star $\iota$~Dra
($V=3.29$, $B-V=1.17$, K2~III). A 2.3 mag neutral-density filter was
used in combination with the $B$ and $V$ filters to attenuate the
signal and so minimize the deadtime correction for the two bright
stars. Three $V-C$ differential magnitudes in both $B$ and $V$ were
computed from each sequence and averaged to create $B$ and $V$ group
means. Group mean differential magnitudes with internal standard
deviations greater than 0.01 mag were rejected to eliminate
observations taken under non-photometric conditions. The surviving
group means were corrected for differential extinction with nightly
extinction coefficients, transformed to the Johnson system with
yearly-mean transformation coefficients, and treated as single
observations thereafter. The typical precision of a single group-mean
observation from T3, as measured for pairs of constant stars, is
$\sim$0.004--0.005 mag \citep{hen00}. The APT acquired one or two
group observations each clear night except for three full nights when
the star was observed at a much higher cadence of $\sim~10$ group
observations per hour. The APT collected a total of 224 $B$ and 220
$V$ group observations. Further details on the automatic operation of
this telescope, the observing procedures, and the data reduction
process can be found in \citet{hen00} and references therein.

\begin{figure}
  \includegraphics[angle=270,width=8.2cm]{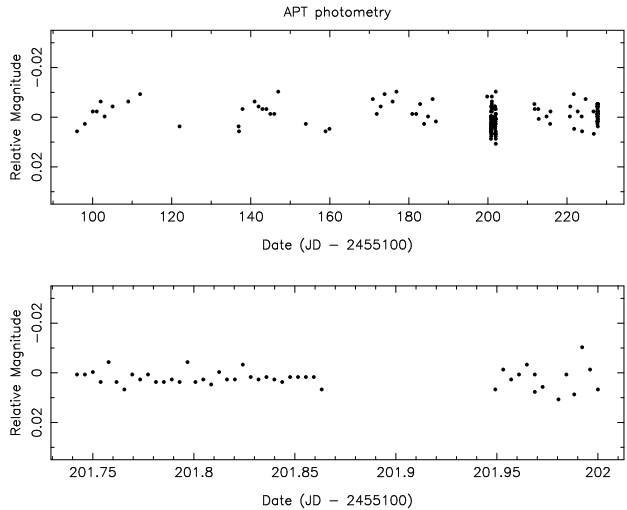}
  \caption{Johnson $B$ photometry of $\iota$ Dra obtained using the T3
    0.4~m APT at Fairborn Observatory showing long-term (top panel) and
    short-term (bottom panel) stability of $\sim0.004$.}
  \label{phot_apt}
\end{figure}

The complete, reduced Johnson $B$ data set is plotted in the top panel
of Figure \ref{phot_apt}; the bottom panel presents just the
high-cadence $B$ photometry from one of the three monitoring nights.
The data in both panels scatter about their means with a standard
deviation of 0.0041 mag, after a half dozen outliers are removed in
each case. Results for the $V$ observations are essentially identical
(0.0043 mag). Thus, our observations suggest that $\iota$~Dra, as
well as its comparison star $theta$~Dra, are both constant to a limit
of approximately 0.004 mag.

\begin{figure}
  \includegraphics[angle=270,width=8.2cm]{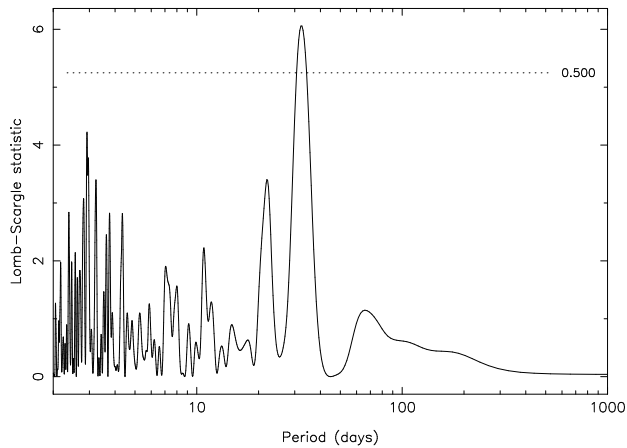}
  \caption{Periodogram of the $B$ observations acquired with the T3 APT.
   No strong periodicity is seen, but the highest peak at 32 days is
   consistent with inspection of the first half of the light curve.}
  \label{phot_pgram}
\end{figure}

We note, however, that the first half of the complete $B$ and $V$ light
curves suggest possible very-low-amplitude variability on a timescale of
$\sim~30$ days. The variation then seems to damp out during the second
half of the light curve. We performed Fourier analyses of the complete
$B$ and $V$ data sets and found no significant periodic variability
in either case (see Figure \ref{phot_pgram} for the $B$ data). These
results are somewhat reminiscent of the light curves and period analyses
of the K giants presented in the survey of \citet{hen00}. Further
observations will be required to determine if the apparent low-amplitude
variability is real. For now, we claim only that $\iota$~Dra is constant
to $\sim~0.004$ mag. We also conclude that the photometric variability
reported by \citet{jac63} is spurious and that $iota$~Dra has yet to prove
its status as a variable star.

In addition to improving the radial velocity orbit of $\iota$~Dra~b and 
detecting a linear trend in the residuals, \citet{zec08} also found 
evidence for short-period radial velocity variations in the range 
0.25--0.33 days (we note that corresponding short-period photometric 
variations are {\it not} seen in our three nights of high-cadence 
observations). Using the radius measured by \citet{all99} of 
$R_\star = 12.88 R_\odot$ and the rotational velocity measured by 
\citet{dem99} of 1.5 km/s, we estimate the expected rotation period 
of $\iota$ Dra to be $\sim 434$ days. Thus, the possible low-amplitude 
variability seen in our photometry and the short-period, low-amplitude 
radial velocity variability found by \citet{zec08} cannot be caused by 
the rotation of the star. \citet{zec08} note that their short-period
radial velocity variations may be similar to solar-type oscillations.  
Likewise, \citet{hen00} concluded that the low-amplitude, short-period 
photometric variability found in their sample of K giants could not be 
due to rotational modulation but had timescales consistent with radial 
pulsation.

%%%%%%%%%%%%%%%%%%%%%%%%%%%%%%%%%%%%%%%%%%%%%%%%%%%%%%%%%%%%%%%%%%%%

\section{Prospects for Transit Detection}

%%%%%%%%%%%%%%%%%%%%%%%%%%%%%%%%%%%%%%%%%%%%%%%%%%%%%%%%%%%%%%%%%%%%

\subsection{Photometric Detection}

The prospects for successful detection of a transit of $\iota$~Dra~b
rely upon achieving the necessary photometric precision. The
brightness of the star makes it a difficult target to observe with
traditional CCD photometry due to saturation. Ground-based
observations with photoelectric photometers, such as the APTs,
mitigate this problem but are still challenged by the $<0.0001$ mag
transit depth. \citet{ass09} suggest observing the star during the
transit window using narrowband filters which would isolate the Ca II H
and K chromospheric emission at the stellar limb, the effect of which
will be to increase the depth of the transit.

While we did not definitely detect photometric variability in
$\iota$~Dra, the amplitude and timescale of possible photometric
variations are significantly larger than the predicted transit
duration. Hence, two possibilities present themselves for detection of
the transit in the face of these potential variations. If the periodic
variation is well-defined over long time periods then it could be
accurately modeled and subtracted from the data. Regardless, the
transit window is shorter such that the variation is not expected to
impact the transit detection on those timescales. Although the long
transit duration prevents the monitoring of a complete transit within
a single night, ground-based observations of a complete transit window
of HD~80606b have been accomplished by \citet{hid10} by obtaining
multi-longitudinal coverage. In addition, the required cadence is much
lower than for a typical hot Jupiter which allows greater flexibility
in binning the measurements for greater precision.

Another possibility is to attempt complete monitoring of the transit
window from a space-based observatory. Consider the case of
HD~209458b, for which the host star is a $V = 7.65$ main sequence
star. Transits of this planet have been observed by both Hubble Space
Telescope (HST) and the Microvariability and Oscillations of Stars
(MOST) satellite at very high precision. The HST observations by
\citet{bro01} achieved a precision of $1.1 \times 10^{-4}$ and the
MOST observations by \citet{cro07} achieved a precision of $3.5 \times
10^{-3}$. Assuming a precision which is dominated by photon noise,
these values convert to $\iota$ Dra precisions of $1.5 \times 10^{-5}$
and $4.7 \times 10^{-4}$ for HST and MOST respectively. These
predicted precisions bring the detection of a transit of $\iota$ Dra b
within reach since the necessarily high cadence of these observations
will require binning of the data over the long transit duration. It
should be noted however that observations of $\iota$ Dra with MOST are
difficult because it is significantly north of the Continous Viewing
Zone (CVZ).

%%%%%%%%%%%%%%%%%%%%%%%%%%%%%%%%%%%%%%%%%%%%%%%%%%%%%%%%%%%%%%%%%%%%

\subsection{Spectroscopic Detection}

The Rossiter-McLaughlin (RM) effect \citep{mcl24,ros24} provides a
means whereby an exoplanet may be detected and indeed discovered
spectroscopically. The first precedent for discovering a planet
through this method occurred with the well-known planet orbiting
HD~189733 \citep{bou05}. The application of the RM effect to
transiting exoplanets has been described in detail by
\citet{gau07}. The amplitude of the RM effect, $K_R$ is
\begin{equation}
  K_R \equiv V_S \sin I_S \frac{\gamma^2}{1-\gamma^2}
\end{equation}
where $V_S \sin I_S$ is the rotational velocity of the star and
$\gamma$ is the transit depth.

The proportionality of the RM amplitude with the rotational velocity
of the star presents an immediate large bias against giant stars due
to their relatively slow rotation rate. For $\iota$ Dra, the
rotational velocity has been measured by \citet{dem99} to be 1.5 km/s.
Using the estimated transit depth of $7.84 \times 10^{-5}$ from Table
\ref{probdepdur}, the amplitude of the RM effect as the planet
transits is predicted to be only 12 cm/s! Any attempt to measure such
a spectroscopic transit signature will undoubtedly require awaiting
the next generation of radial velocity instruments.

%%%%%%%%%%%%%%%%%%%%%%%%%%%%%%%%%%%%%%%%%%%%%%%%%%%%%%%%%%%%%%%%%%%%

\section{Conclusions}

Inadequate precision of orbital parameters presents a major hinderance
to testing the transit hypothesis of planets discovered using the
radial velocity technique. The challenge of improving transit
ephemerides and monitoring transit windows is currently being met by
such groups as the Transit Ephemeris Refinement and Monitoring Survey
(TERMS). The value of such a survey includes an improved understanding
of planetary orbits and the potential for gaining insight into the
exoplanet mass-radius relationship for the brightest host stars.

For giant stars, this regime is largely unexplored even though giant
stars have the highest transit probabilities since $P_t \propto
R_\star$. Eccentric orbits of planets around giant stars are
relatively rare, with only a handful exceeding an eccentricity of 0.5,
including HD~122430b and $\iota$ Dra b. The planet orbiting HD~13189
\citep{hat05} has an eccentricity of 0.28 but this is more than
compensated for by the host; a K2 II star with an estimated radius
exceeding 50 $R_\sun$.

The case of $\iota$ Dra b is particularly interesting since the
stellar radius and orbital elements create a transit probability
exceeding that of a hot Jupiter ($P < 5$ days), despite having an
orbital period of more than 500 days. Our refined orbital solution
produces a transit window that is accessible to a concerted
ground-based effort or modest space-based effort with potentially high
return for a successful transit detection. In addition, our photometry
shows the short-term stability of $\iota$ Dra to be comparatively
favourable for typical late-type giant stars. A fourier analysis of
the photometry reveals a periodic signal at $\sim 32$ days, although
the weakness of the signal makes its presence inconclusive and we
conclude that $\iota$~Dra is constant at the $\sim 0.004$ mag level.
However, the challenge of achieving the needed photometric precision
required to achieve an unambiguous detection for the transit will
likely require spaced-based observations during the transit window.

%%%%%%%%%%%%%%%%%%%%%%%%%%%%%%%%%%%%%%%%%%%%%%%%%%%%%%%%%%%%%%%%%%%%

\section*{Acknowledgements}

The authors would like to thank David Ciardi and Andrew Howard for
several useful suggestions. We thank the many CAT observers who have
observed $\iota$ Dra over the years, especially David Mitchell and Saskia
Hekker. Contributions of Andreas Quirrenbach to the K giants radial
velocity monitoring program at Lick Observatory are gratefully
acknowledged. This research has made use of the NASA/IPAC/NExScI Star
and Exoplanet Database, which is operated by the Jet Propulsion
Laboratory, California Institute of Technology, under contract with
the National Aeronautics and Space Administration.

%%%%%%%%%%%%%%%%%%%%%%%%%%%%%%%%%%%%%%%%%%%%%%%%%%%%%%%%%%%%%%%%%%%%


\begin{thebibliography}{}

\bibitem[\protect\citeauthoryear{Allende Prieto \&
    Lambert}{1999}]{all99} Allende Prieto, C., Lambert, D.L., 1999,
  A\&A, 352, 555
\bibitem[\protect\citeauthoryear{Assef et al.}{2009}]{ass09}
  Assef, R.J., Gaudi, B.S., Stanek, K.Z., 2009, ApJ, 701, 1616
\bibitem[\protect\citeauthoryear{Barbieri et al.}{2007}]{bar07a}
  Barbieri, M., et al., 2007, A\&A, 476, L13
\bibitem[\protect\citeauthoryear{Barnes}{2007}]{bar07b} Barnes, J.W.,
  2007, PASP, 119, 986
\bibitem[\protect\citeauthoryear{Batygin et al.}{2009}]{bat09}
  Batygin, K., Bodenheimer, P., Laughlin, G., 2009, ApJ, 704, L49
\bibitem[\protect\citeauthoryear{Bodenheimer et al.}{2003}]{bod03}
  Bodenheimer, P., Laughlin, G., Lin, D.N.C., 2003, ApJ, 592, 555
\bibitem[\protect\citeauthoryear{Bouchy et al.}{2005}]{bou05} Bouchy,
  F., et al., 2005, A\&A, 444, L15
\bibitem[\protect\citeauthoryear{Brown et al.}{2001}]{bro01} Brown,
  T.M., Charbonneau, D., Gilliland, R.L., Noyes, R.W., Burrows, A.,
  2001, ApJ, 552, 699
\bibitem[\protect\citeauthoryear{Croll et al.}{2007}]{cro07}
  Croll, B., et al., 2007, ApJ, 658, 1328
\bibitem[\protect\citeauthoryear{da Silva et al.}{2006}]{das06}
  da Silva, L., et al., 2006, A\&A, 458, 609
\bibitem[\protect\citeauthoryear{de Medeiros \& Mayor}{1999}]{dem99}
  de Medeiros, J.R., Mayor, M., 1999, A\&AS, 139, 433
\bibitem[\protect\citeauthoryear{D\"ollinger et al.}{2009}]{dol09}
  D\"ollinger, M.P., Hatzes, A.P., Pasquini, L., Guenther, E.W.,
  Hartmann, M., 2009, A\&A, 505, 1311
\bibitem[\protect\citeauthoryear{Frink et al.}{2002}]{fri02} Frink,
  S., Mitchell, D.S., Quirrenbach, A., Fischer, D.A., Marcy, G.W.,
  Butler, R.P., 2002, ApJ, 576, 478
\bibitem[\protect\citeauthoryear{Gaudi \& Winn}{2007}]{gau07} Gaudi,
  B.S., Winn, J.N., 2007, ApJ, 655, 550
\bibitem[\protect\citeauthoryear{Hatzes et al.}{2005}]{hat05} Hatzes,
  A.P., Guenther, E.W., Endl, M., Cochran, W.D., D\"ollinger, M.P.,
  Bedalov, A., 2005, A\&A, 437, 743
\bibitem[\protect\citeauthoryear{Henry et al.}{2000}]{hen00} Henry,
  G.W., Fekel, F.C., Henry, S.M., Hall, D.S., 2000, ApJS, 130, 201
\bibitem[\protect\citeauthoryear{Hidas et al.}{2010}]{hid10}
  Hidas, M.G., et al., 2010, MNRAS, 406, 1146
\bibitem[\protect\citeauthoryear{Hoffleit \& Jaschek}{1991}]{hof91}
  Hoffleit, D., Jaschek, C.V., 1991, The Bright Star Catalogue, 5th
  edition (New Haven: Yale Univ. Obs.)
\bibitem[\protect\citeauthoryear{Ida \& Lin}{2005}]{ida05} Ida, S.,
  Lin, D.N.C., 2005, ApJ, 626, 1045
\bibitem[\protect\citeauthoryear{Jackisch}{1963}]{jac63} 
  Jackisch, G., 1963, Veroeffentlichungen der Sternwarte in Sonneberg
  (Publications of Sonneberg Observatory), Bd. 5, Heft, 5., p. 232
\bibitem[\protect\citeauthoryear{Kane \& von Braun}{2008}]{kan08}
  Kane, S.R., von Braun, K., 2008, ApJ, 689, 492
\bibitem[\protect\citeauthoryear{Kane \& von Braun}{2009}]{kan09a}
  Kane, S.R., von Braun, K., 2009, PASP, 121, 1096
\bibitem[\protect\citeauthoryear{Kane et al.}{2009}]{kan09b} Kane,
  S.R., Mahadevan, S., von Braun, K., Laughlin, G., Ciardi, D.R.,
  2009, PASP, 121, 1386
\bibitem[\protect\citeauthoryear{Kennedy \& Kenyon}{2008}]{ken08}
  Kennedy, G.M., Kenyon, S.J., 2008, ApJ, 673, 502
\bibitem[\protect\citeauthoryear{Kukarkin et al.}{1982}]{kuk82}
  Kukarkin, B.V., et al., 1982, New Catalogue of Suspected Variable
  Stars, Moscow: Academy of Sciences
\bibitem[\protect\citeauthoryear{Laughlin et al.}{2009}]{lau09}
  Laughlin, G., Deming, D., Langton, J., Kasen, D., Vogt, S., Butler,
  P., Rivera, E., Meschiari, S., 2009, Nature, 457, 562
\bibitem[\protect\citeauthoryear{McLaughlin}{1924}]{mcl24} McLaughlin,
  P.R., 1924, ApJ, 60, 22
\bibitem[\protect\citeauthoryear{Moutou et al.}{2009}]{mou09} Moutou
  et al., 2009, A\&A, 498, L5
\bibitem[\protect\citeauthoryear{Niedzielski et al.}{2009}]{nie09}
  Niedzielski, A., Go\'zdziewski, K., Wolszczan, A., Konacki, M.,
  Nowak, G., Zieli\'nski, P., 2009, ApJ, 693, 276
\bibitem[\protect\citeauthoryear{Percy}{1993}]{per93}
  Percy, J.R., 1993, PASP, 105, 1422
\bibitem[\protect\citeauthoryear{Perryman et al.}{1997}]{per97}
  Perryman, M.A.C., et al., 1997, A\&A, 323, L49
\bibitem[\protect\citeauthoryear{Rossiter}{1924}]{ros24} Rossiter,
  R.A., 1924, ApJ, 60, 15
\bibitem[\protect\citeauthoryear{Sadakane et al.}{2005}]{sad05}
  Sadakane, K., Ohnishi, T., Ohkubo, M., Takeda, Y., 2005, PASJ, 57,
  127
\bibitem[\protect\citeauthoryear{Setiawan et al.}{2004}]{set04}
  Setiawan, J., Pasquini, L., da Silva, L., Hatzes, A.P., von der
  L\"uhe, O., Girardi, L., de Medeiros, J.R., Guenther, E., 2004,
  A\&A, 421, 241
\bibitem[\protect\citeauthoryear{Veras et al.}{2009}]{ver09} Veras,
  D., Crepp, J.R., Ford, E.B., 2009, ApJ, 696, 1600
\bibitem[\protect\citeauthoryear{Wright}{2005}]{wri05} Wright, J.T.,
  2005, PASP, 117, 657
\bibitem[\protect\citeauthoryear{Zechmeister et al.}{2008}]{zec08}
  Zechmeister, M., Reffert, S., Hatzes, A.P., Endl, M., Quirrenbach,
  A., 2008, A\&A, 491, 531

\end{thebibliography}
\end{document}